\documentclass{article}
\PassOptionsToPackage{numbers, compress}{natbib}

\usepackage[final]{neurips_2020}

\usepackage[utf8]{inputenc} 
\usepackage[T1]{fontenc}    
\usepackage{hyperref}       
\usepackage{url}            
\usepackage{booktabs}       
\usepackage{amsfonts}       
\usepackage{nicefrac}       
\usepackage{microtype}      
\usepackage{graphicx}
\usepackage[english]{babel}
\usepackage{amssymb,amsmath,amsthm}
\usepackage{thmtools}
\usepackage{thm-restate}
\usepackage{cleveref}

\allowdisplaybreaks
\title{Self-Supervised Out-of-Distribution Detection \\in Brain CT Scans}

\usepackage{verbatim} 

\author{Abinav Ravi Venkatakrishnan\thanks{First two authors contributed equality to this work.} $^{,1}$, \textbf{Seong Tae Kim}$^{*,}$\thanks{Corresponding author (seongtae.kim@tum.de).}\;$^{,1}$, \textbf{Rami Eisawy}$^{1,2}$, \\
\textbf{Franz Pfister}$^{2,3}$, \textbf{Nassir Navab}$^{1,4}$\\[1ex]
$^1$Technical University of Munich, $^2$DeepC GmbH \\
$^3$Ludwig Maximilian University of Munich, $^4$Johns Hopkins University
}
\begin{document}

\maketitle
\begin{abstract}
Medical imaging data suffers from the limited availability of annotation because annotating 3D medical data is a time-consuming and expensive task. Moreover, even if the annotation is available, supervised learning-based approaches suffer highly imbalanced data. Most of the scans during the screening are from normal subjects, but there are also large variations in abnormal cases. To address these issues, recently, unsupervised deep anomaly detection methods that train the model on large-sized normal scans and detect abnormal scans by calculating reconstruction error have been reported. In this paper, we propose a novel self-supervised learning technique for anomaly detection. Our architecture largely consists of two parts: 1) Reconstruction and 2) predicting geometric transformations. By training the network to predict geometric transformations, the model could learn better image features and distribution of normal scans. In the test time, the geometric transformation predictor can assign the anomaly score by calculating the error between geometric transformation and prediction. Moreover, we further use self-supervised learning with context restoration for pretraining our model. By comparative experiments on clinical brain CT scans, the effectiveness of the proposed method has been verified.
\end{abstract}

\section{Introduction}

Supervised deep learning methods have achieved state-of-the-art performances various tasks \cite{image_classification}.
However, medical imaging data suffer from the limited availability of annotation because annotating 3D medical data is a time-consuming and expensive task. Moreover, even if the annotation is available, supervised learning-based approaches suffer highly imbalanced data. In other words, most of the scans during the screening are from normal subjects, and only a few subjects are the abnormal with a large intraclass variation. 
To address these issues, recently, unsupervised deep anomaly detection methods \cite{anovaegan,cevae, context-restoration, f-Anogan, schlegl2017unsupervised, VAE-KL, UAD-review, zimmerer2019high} have been introduced based on autoencoder-based reconstruction methods \cite{AE} or variational autoencoders (VAE) \cite{VAE_orig}. These methods train a model on large-sized normal scans and detect abnormal scans by calculating reconstruction error because abnormal scans are out-of-distribution, making the model have difficulties reconstructing abnormal areas.

In this paper, we propose a novel anomaly detection method which based on self-supervised learning \cite{ood_self_supervised,rotnet}. Our architecture largely consists of two parts: 1) Reconstruction and 2) predicting geometric transformations. By training the network to predict geometric transformations, the model could learn the distribution of normal scans in a better way. In the test time, the geometric transformation predictor can assign the anomaly score by calculating the error between the geometric transformation we apply for the given sample and the prediction of the deep neural network for the geometric transformations.
Moreover, we further use self-supervised learning with context restoration for pretraining our model by learning better image features from complex medical image \cite{context-restoration}. 

Our main contributions are: (1) We introduce a new anomaly detection framework that defines an anomaly score by considering both pixel-level reconstruction and image-level geometric transformation prediction. The reconstruction-based module exploits the local fine-grained information to detect out-of-distribution pixels. The image-level geometric transformation module captures the global context to define an anomaly score.  
(2) To verify the effectiveness of our method, we collected clinical data where we have various anomaly cases in the test set. By comparative experiments on clinical brain CT scans, we show that our method can detect anomaly samples more accurately because it utilizes the complementary information of the local reconstruction-based approach and global geometric transformation-based approach.

\begin{figure}
    \centering
    \includegraphics[width=0.95\textwidth, scale=2.0]{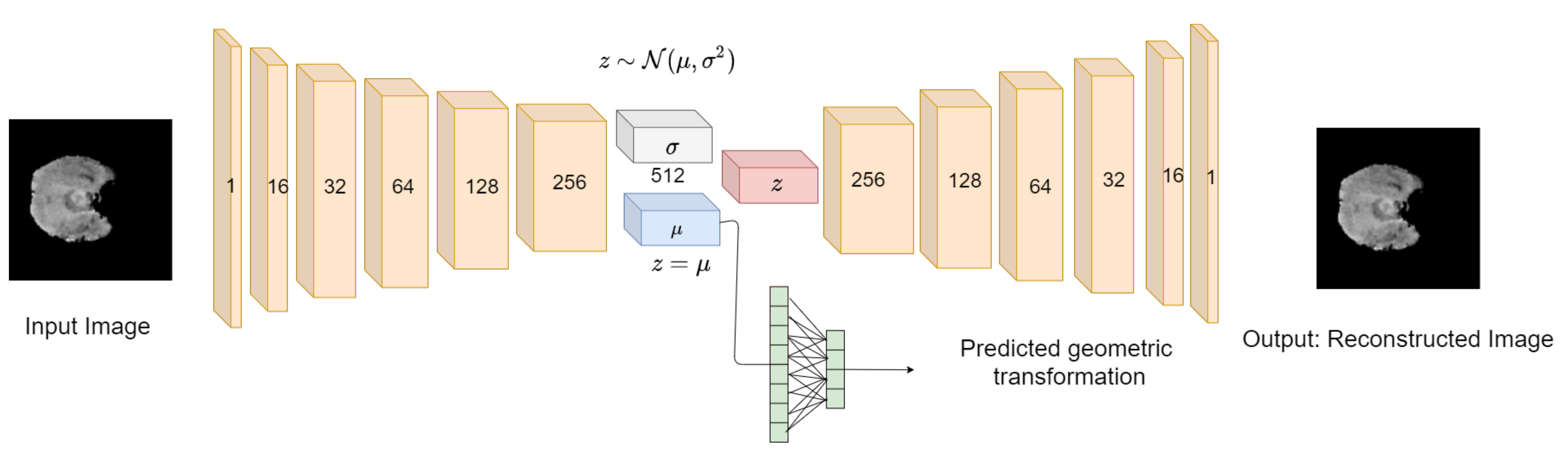}
    \caption{Proposed network architecture for anomaly detection. The numbers in the Blocks denote the number of filters in the convolution layer. $z$ represents the latent space, $ \sigma $ and $ \mu $ represent the variance and mean of the latent distribution.  There are two outputs for the model: 1) reconstructed images, 2) predicted geometric transformation.} 
    \label{fig:Multi_task_architecture}
\end{figure}

\section{Method and Materials}
We propose a framework for predicting the anomaly score based on global geometric information and local fine-grained information. 

As shown in Figure 1, our framework consists of VAE for reconstructing the image based on distribution learned during the training and geometric transformation predictor for estimating the geometric transformation applied for the given input image.
The details will be introduced in the following subsections.

\label{sec:pretrain}
\textbf{Pretraining with Context Restoration. }To exploit the information in the training data and learn general features of brain CT images, we first train our model with the context restoration. The VAE structure is trained by restoring the context from inputs whose patches are swapped. For swapping the patches in the image, two points are extracted from non-zero parts of the image in axial orientation, and then a patch of size $\frac{1}{8}\times$ image size is created with the extracted point at the center of the patch. These patches are then swapped. The objective function is designed to reconstruct the original image from the swapped image as 

\begin{equation}
\label{eqn:reconstruction loss}
    \mathcal{L}_{cr} = \| (x_i - f(\hat{x}_i))\|_2
\end{equation}

where $x_i$, $\hat{x}_i$ denote the original input image and the swapped image, respectively. $f$ denotes the function for variational autoencoder.

\label{sec:SSl}
\textbf{Multi Task Learning. }As shown in Figure 1, our model consists of VAE and geometric transformation predictor. 
Inspired by \cite{ad-geometric-transforms,ood_self_supervised}, the self-supervised learning based on pretext from geometric transformation is used in our method.
For training geometric transformation predictor, the labels $q_i$ are generated by rotation the sample in $[0^{\circ},90^{\circ},180^{\circ},270^{\circ}$] at random or translating the image by $\frac{1}{8}$th of the image size in vertical or horizontal. The number of permutations of rotations and translation combinations gives the number of neurons at the output of the fully connected layer. The rotation and translation help in learning the global geometric features of in-distribution healthy scans. The geometric transformation predictor is trained by 
\begin{equation}
\label{classification loss}
    \mathcal{L}_{geo} = -\sum_{i}^{N} q_i \log g(\tilde{x}_{q_i})
\end{equation}

where $g$ denotes the function for geometric transformation predictor which produces the output for representing the probability for the corresponding transformation, $\tilde{x}_{q_i}$ denotes the geometric transformed input image by ${q_i}$.

In addition to the geometric transformation predictors, the VAE is fine-tuned to reconstruct the transformed images. As a result, the overall loss function for our model is defined as

\begin{equation}
\label{eq:Multi task loss}
    \mathcal{L}_{multitask} = \mathcal{L}_{geo} + \epsilon \mathcal{L}_{rec} 
\end{equation}

where $\mathcal{L}_{rec}= \| (x_i - f({x}_i))\|_2$ 
and $\epsilon$ denotes a scaling factor that balance the two loss terms. 

\textbf{Anomaly Score Calculation. }The proposed method has two different outputs, one corresponding to the softmax score from the geometric transformation predictor, and the second output is the reconstruction of the images from the decoder part. The scores from both the outputs are combined to give a single anomaly score.  The score is calculated using the normalized weighted averaging using parameter $\lambda$ which serves as a hyperparameter. The combination can be represented as follows. 

\begin{equation}
    score = (1 - \lambda)s_g + \lambda s_r 
\end{equation}

where $s_g$ and $s_r$ represents the anomaly scores from the geometric transformation head and reconstruction, respectively. The anomaly score from the geometric transformation $s_g$ is averaged value obtained from possible variations in the geometric transformation in the test time. 
The anomaly score from reconstruction was defined as $s_r=\alpha\times\| (x_i - f({x}_i))\|_2$. $\alpha$ is the scaling factor. The $\lambda$ value is empirically selected to be 0.5 in the experiments. 

\textbf{Dataset.} To evaluate our approach, the clinical brain CT scans are used. Table 1 shows detailed information for the dataset. In particular, the test set contains various anomalies as followings: atrophy (20 volumes), intracranial bleeding (11 volumes), ischemia (9 volumes), cavernoma (1 volume), aneurysm (1 volume), bleed (1 volume), and tumor (1 volume). Please note that the models are trained from only normal slices.

\begin{table}[t]
\begin{center}

     {\begin{tabular}{|c | c | c | c | c |} 
     \hline
    & Normal volumes & Normal slices & Abnormal volumes & Abnormal slices\\ 
     \hline
     Training & 119 & 16506 & - & - \\ 
     \hline
     Validation & 30 & 4126 &- &-  \\
     \hline
     Test & 30 & 4314 & 44 & 4805 \\
     \hline
    \end{tabular}}
 \label{tab:data}
 \caption{Distribution of the number of CT volumes and slices in this study.}
\end{center}
\end{table}

\section{Results and Discussion}
\label{sec:metrics}
\textbf{Metrics. }To measure the anomaly detection performance (classification of anomaly slice and normal slice), Area under the receiver operating characteristic curve (AUROC) and Area under the precision-recall curve (AUPR) are adopted. In addition, we also provide the segmentation performance based on the dice similarity coefficient (DSC) because previous anomaly detection approaches evaluated the segmentation performance \cite{cevae, VAE-KL}. Please note that our method more focuses on anomaly detection, not segmentation.

\textbf{Results.} For comparison, we implemented state-of-the-art anomaly detection methods in brain images \cite{cevae, VAE-KL}. For the ablation study, we also implement the model with only reconstruction loss $L_{rec}$ after pretraining (i.e., VAE with Pretraining (ours)).
Table 2 shows the results of the anomaly detection and segmentation. As shown in the table, by encouraging the model to learn better medical features with pretraining (context restoration), VAE with Pretraining improves AUROC and AUPR compared with VAE \cite{VAE-KL}. Moreover, our multitask framework further improves the classification performance in terms of AUROC and AUPR and it outperforms \cite{cevae, VAE-KL}. However, the localization performance of the multi-task framework is slightly worse than others because the encoder is jointly trained to reconstruct images and predict at the same time and the segmentation performance fully depends on the reconstruction part.

\begin{table}[t]
\begin{center}
 \begin{tabular}{|c | c | c | c |} 
 \hline
& AUROC & AUPR  & DSC\\ 

 \hline
 VAE \cite{VAE-KL} & 0.668 & 0.704 &  0.110 $\pm$ 0.021\\ 
 \hline
 Context-encoding VAE \cite{cevae} & 0.766 & 0.640 & \textbf{0.112 $\pm$ 0.021 } \\
 \hline

 \textbf{VAE with Pretraining (ours) }& 0.673 & 0.772  & \textbf{0.112 $\pm$ 0.021 }\\
 \hline
 \textbf{Multi-task Framework (ours)}  & \textbf{0.822} & \textbf{0.868}  & 0.086 $\pm$ 0.024\\ [1ex] 
 \hline
\end{tabular}
\label{ tab:AUPR}
 \caption{Comparison with state-of-the-art anomaly detection methods.}
\end{center}
\end{table}

\textbf{Discussion and Conclusion. }This paper introduces a novel architecture for anomaly detection in brain CT scans. 
Our method uses two-types of self-supervision to consider the global context to detect anomaly images and also improve the model in the pretraining stage.
The experimental results on clinical brain CT images show that the pretraining with self-supervised learning (context restoration) can improve the performances in terms of classification and localization.
Moreover, the proposed multi-task framework outperforms other anomaly detection methods in terms of anomaly detection. 
However, there is a tradeoff between the classification performance and the segmentation performance. It should be deployed depending on the real-world application (i.e., whether the classification is important or the localization is important). This paper verifies the importance of self-supervised learning in anomaly detection problems and can present a new direction for future research in the medical domain.

\section*{Acknowledgements}
The authors are thankful to DeepC Gmbh for their help with the dataset and the computing resources during this project.

\bibliographystyle{unsrt}
\bibliography{neurips_2020}

\begin{thebibliography}{10}

\bibitem{image_classification}
Alex Krizhevsky, Ilya Sutskever, and Geoffrey~E Hinton.
\newblock Imagenet classification with deep convolutional neural networks.
\newblock {\em Advances in Neural Information Processing Systems}, pages
  1097--1105, 2012.

\bibitem{anovaegan}
Christoph Baur, Benedikt Wiestler, Shadi Albarqouni, and Nassir Navab.
\newblock Deep autoencoding models for unsupervised anomaly segmentation in
  brain mr images.
\newblock {\em International MICCAI Brainlesion Workshop}, 2018.

\bibitem{cevae}
David Zimmerer, Jens Petersen, Fabian Isensee, and Klaus Maier-Hein.
\newblock Context-encoding variational autoencoder for unsupervised anomaly
  detection.
\newblock {\em International Conference on Medical Imaging with Deep Learning},
  2019.

\bibitem{context-restoration}
Liang Chen, Paul Bentley, Kensaku Mori, Kazunari Misawa, Michitaka Fujiwara,
  and Daniel Rueckert.
\newblock Self-supervised learning for medical image analysis using image
  context restoration.
\newblock {\em Medical Image Analysis}, 58:101539, 2019.

\bibitem{f-Anogan}
Thomas Schlegl, Philipp Seeböck, Sebastian~M. Waldstein, Georg Langs, and
  Ursula Schmidt-Erfurth.
\newblock f-anogan: Fast unsupervised anomaly detection with generative
  adversarial networks.
\newblock {\em Medical Image Analysis}, 54:30 -- 44, 2019.

\bibitem{schlegl2017unsupervised}
Thomas Schlegl, Philipp Seeb{\"o}ck, Sebastian~M Waldstein, Ursula
  Schmidt-Erfurth, and Georg Langs.
\newblock Unsupervised anomaly detection with generative adversarial networks
  to guide marker discovery.
\newblock {\em International Conference on Information Processing in Medical
  Imaging}, pages 146--157, 2017.

\bibitem{VAE-KL}
David Zimmerer, Fabian Isensee, Jens Petersen, Simon Kohl, and Klaus
  Maier-Hein.
\newblock Unsupervised anomaly localization using variational auto-encoders.
\newblock {\em International Conference on Medical Image Computing and
  Computer-Assisted Intervention}, pages 289--297, 2019.

\bibitem{UAD-review}
Christoph Baur, Stefan Denner, Benedikt Wiestler, Shadi Albarqouni, and Nassir
  Navab.
\newblock Autoencoders for unsupervised anomaly segmentation in brain mr
  images: A comparative study.
\newblock {\em arXiv preprint arXiv:2004.03271}, 2020.

\bibitem{zimmerer2019high}
David Zimmerer, Jens Petersen, and Klaus Maier-Hein.
\newblock High-and low-level image component decomposition using vaes for
  improved reconstruction and anomaly detection.
\newblock {\em Medical Imaging Meets NeurIPS}, 2019.

\bibitem{AE}
David~E Rumelhart, Geoffrey~E Hinton, and Ronald~J Williams.
\newblock Learning internal representations by error propagation.
\newblock {\em California Univ San Diego La Jolla Inst for Cognitive Science},
  1985.

\bibitem{VAE_orig}
Diederik~P. Kingma and Max Welling.
\newblock Auto-encoding variational bayes.
\newblock {\em International Conference on Learning Representations (ICLR)},
  2014.

\bibitem{ood_self_supervised}
Dan Hendrycks, Mantas Mazeika, Saurav Kadavath, and Dawn Song.
\newblock Using self-supervised learning can improve model robustness and
  uncertainty.
\newblock {\em Advances in Neural Information Processing Systems (NeurIPS)},
  pages 15637--15648, 2019.

\bibitem{rotnet}
Spyros Gidaris, Praveer Singh, and Nikos Komodakis.
\newblock Unsupervised representation learning by predicting image rotations.
\newblock {\em International Conference on Learning Representations (ICLR)},
  2018.

\bibitem{ad-geometric-transforms}
Izhak Golan and Ran El-Yaniv.
\newblock Deep anomaly detection using geometric transformations.
\newblock {\em Advances in Neural Information Processing Systems}, pages
  9758--9769, 2018.

\bibitem{simclr}
Ting Chen, Simon Kornblith, Mohammad Norouzi, and Geoffrey Hinton.
\newblock A simple framework for contrastive learning of visual
  representations.
\newblock {\em ICML}, 2020.

\end{thebibliography}

\section{A Broader Impact Statement}
Deep learning in Computer-aided diagnosis has helped in enhancing the life of clinical practitioners and patients. To deploy deep learning algorithms we need annotations for training the model which are expensive in medical applications. Self-supervised learning tries to solve the problem by using self-annotated data and obtains a very good performance compared to supervised learning tasks \cite{simclr}. This work will help progress the field of anomaly detection using self-supervised learning. The problem of unsupervised anomaly detection has always been considered from a segmentation perspective. In this work, we try to extend this work to the classification regime by including self-supervised out of distribution detection. This paper verifies the importance of self-supervised learning in anomaly detection problems and presents a new direction for future research in the medical domain.

\end {document}